# Up to 168-Gb/s PAM-4 direct detection transmission with silicon travelling wave Mach-Zehnder modulator

Fan Yang, Yixiao Zhu, Lei Zhang, Xiaoke Ruan, Yanping Li and Fan Zhang
*State Key Laboratory of Advanced Optical Communication System and Networks, Peking University*
Beijing, China
fzhang@pku.edu.cn

*Abstract*—Based on a silicon travelling wave Mach-Zehnder modulator, we demonstrate record 168Gb/s (84Gbaud) and 160Gb/s (80Gbaud) PAM-4 direct detection transmission over 1km and 2km SSMF, respectively, with bit error rates below 20% HD-FEC threshold.

## I. Introduction

With the growing demand for broadband services, optical transceivers with lane rate beyond 100G is highly desired for the data center interconnections. For short reach transmission systems with distance below 2km, intensity modulation (IM) with direct detection (DD) is the most promising solution because of its low cost and easy integration [1]. Compared with half-cycle subcarrier modulation (SCM), multi-band carrier-less amplitude phase modulation (CAP), and discrete multi-tone modulation (DMT), pulse amplitude modulation (PAM) has the advantage of simpler structure. In [2], based on a C-band silicon single-drive traveling wave (TW) Mach-Zehnder modulator (MZM), 112Gb/s PAM-4 signal is generated and transmitted over 2km standard single mode fiber (SSMF) with a bit error rate (BER) below 7% hard-decision forward error-correction (HD-FEC) threshold of $3.8\times10^{-3}$. With the same TW-MZM, a 128Gb/s PAM-4 signal is transmitted over 1km SSMF with a BER below $3.8\times10^{-3}$. Using a multi electrode MZM (ME-MZM), 168Gb/s PAM-4 is achieved with a BER below $2.1\times10^{-4}$ [2]. Recently, with a substrate removed TW-MZM, 128Gb/s PAM-4 signal is transmitted over 2km SSMF with a BER below $2\times10^{-2}$ [3]. Although ME-MZM and substrate removed TW-MZM can extend the modulator bandwidth significantly [2, 3], both of them have inherent disadvantages. For ME-MZM, the phase alignment and time skew between different segments requires finely adjusting. The group delay variation of electrical amplifier for each segment should be a limitation for practical application in high baud rate. The substrate removed TW-MZM will increase fabrication difficulty and cost. Therefore, to meet the low-cost requirement in data-center applications, high baud rate operation of conventional silicon TW-MZM is of great importance due to its simple structure.

In this work, we experimentally demonstrate the generation and transmission of high baud rate Nyquist PAM-4 signal by using a silicon dual-drive TW-MZM with a bandwidth of ~21GHz. To be specific, by using post filter and maximum likelihood sequence detection (MLSD), we achieve 168Gb/s (84Gbaud) and 160Gb/s (80Gbaud) PAM-4 signal transmission over 1km and 2km SSMF, respectively, with a BER below the 20% HD-FEC threshold of $1.5\times10^{-2}$ [5]. To the best of our knowledge, our work reports the highest single-lane bit rate for direct detection transmission with a silicon TW-MZM.

## II. Characterization of Silicon TW-MZM

Fig. 1(a) shows the structure of the silicon TW-MZM fabricated through IMEC's silicon photonics ISIPP50G technology. Each arm of the TWMZM contains a phase modulator of 1.5 mm length. The phase modulators operate via the plasma dispersion effect, where the depletion of free carriers from a reverse biased PN junction embedded in the waveguide causes a phase shift of the propagating light. Each signal line is terminated with a 25Ω on-chip resistor, established by two parallel 50Ω n-doped silicon slabs between the ground and the signal. On the left side of the signal lines, a group of GSGSG pads are built for electrical RF probe to apply the bias voltage and the high-speed driving signal. The waveguides of the two arms are intentionally designed with 20μm length difference outside the modulation region, which allows the modulator's operation point to be adjusted through

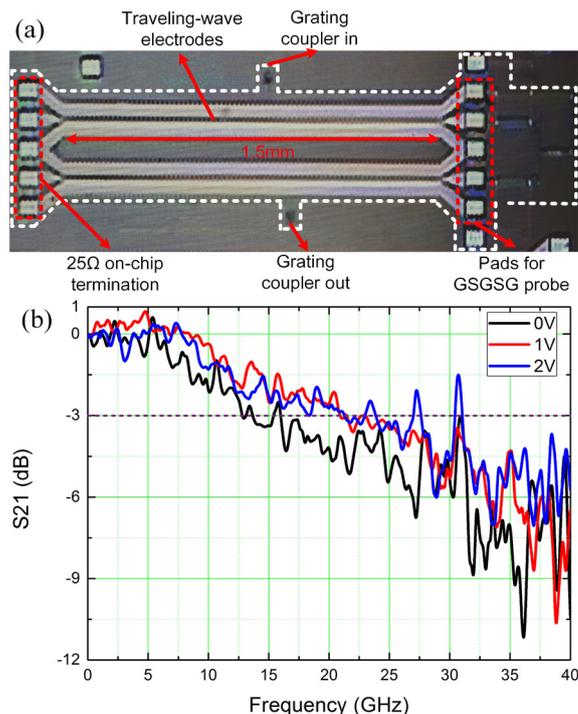

Fig. 1. (a) Micrograph of the TW-MZM. (b) Electro-optic response of the TW-MZM.

This work was supported by National Natural Science Foundation of China (61475004 and 61535002). The authors would like to thank Keysight Technologies for the loan of AWG.

wavelength tuning. Light is coupled in and out of the waveguide device by fiber-to-chip grating couplers with an insertion loss of ~5 dB/coupler. Fig. 1(b) shows the measured electro-optic response of the TW-MZM under different reverse bias voltages. By changing the reverse bias voltage from 0 V to 2 V, the 3-dB bandwidth of the TW-MZM gradually increases from ~13 GHz to ~21 GHz and then stays almost as a constant. Here we chose 1V as the reverse bias in the following experiments.

## III. EXPERIMENTAL SETUP AND DSP STACK

The experimental setup and DSP stack are shown in Fig.2. At the transmitter, an external cavity lasers (ECL) with ~100kHz linewidth is employed as the optical source. A polarization controller (PC) is used to adjust the polarization state of the light coupling to the chip. The baseband Nyquist PAM-4 signal is generated by an arbitrary waveform generator (AWG) (Keysight M8196A) operating at 92GSa/s. After amplified by a pair of electrical amplifiers (EAs) with 50GHz bandwidth, the differential electrical waveforms from AWG are used to drive the TW-MZM through a GSGSG microwave probe with 40GHz bandwidth. For linear intensity modulation, the TW-MZM is biased at the quadrature point by tuning the wavelength of ECL to 1545.72nm. After 1 (or 2) km SSMF transmission, a variable optical attenuator (VOA) and an erbium-doped fiber amplifier (EDFA) are employed at the receiver to control the received power and loading amplified spontaneous emission (ASE) noise. Afterwards the signal is detected with a pin photodiode (PD) and amplified by an electrical amplifier (EA). The bandwidth of PD and EA are both 50GHz. Finally, the electrical waveform is captured by a real-time digital storage oscilloscope (DSO) (Keysight DSA-X 96204Q) with sample rate of 160GSa/s for offline processing.

Fig.2(b) shows the diagram of the DSP stack. At the transmitter, the binary bit stream is mapped to PAM-4 symbol first. Taking 80Gbaud signal generation as an example, the data symbols are 23-times up-sampled before digital shaping with root raise cosine (RRC) filter. The roll-off factor of RRC filter is set as 0.01 to construct a rectangular-like signal spectrum. The preamble includes two 64-symbol M-sequences for synchronization and four 128-symbol M-sequences for channel estimation and equalization. 20000 data symbols are transmitted after the preamble. Therefore, the net bit rate of a single channel is 129.2(=80×2/1.2/(64×2+128×4+20000) ×20000) Gb/s with consideration of both frame redundancy and the 20% overhead HD-FEC. Finally, the signal is sent to the AWG after 20-times down-sampling.

At the receiver, the captured waveform is firstly re-sampled to 2 samples per symbol (SPS). After matched RRC filter, the signal is synchronized according to the cross correlation of the M-sequence. Linear channel equalization is performed in the time domain. The channel response is estimated based on the training sequences, in which the filter taps are updated according to the recursive least square (RLS) algorithm for fast convergence. Then the equalizer is convoluted with data symbols. Afterwards, a decision-directed RLS filter is used to mitigate the residual inter-symbol interference (ISI) by tracing the time-varying channel response. Note that our system is bandwidth-limited for such high baud rate signal transmission. Therefore, linear equalization would enhance the in-band high frequency noise. Here we apply the post equalization method, in which the signal is first digitally convoluted with a two-tap post filter, and subsequently MLSD is used to eliminate the ISI [5]. Finally, the BER is calculated by error counting based on a total of $2\times10^5$ bit samples.

## IV. RESULTS AND DISCUSSION

Fig.3 shows the measured BER versus alpha in post filter for 80/84Gbaud PAM-4 signal after different transmission distance, respectively. Larger alpha achieves better performance for higher baud rate or longer transmission distance. For back-to-back (BTB) and 1km SSMF transmission, the optimal values of the tap alpha are 0.5 and 0.6, respectively, for 84Gbaud and 80Gbaud signals. For 2km SSMF transmission, the optimal alpha values are 0.8 and 0.9 for 80Gbaud and 84Gbaud signals, respectively.

Fig.4(a)-(c) display the measured BERs as a function of bit rate with/without post filter and MLSD at BTB scenario, after 1km and 2km SSMF transmission, respectively. For the BTB

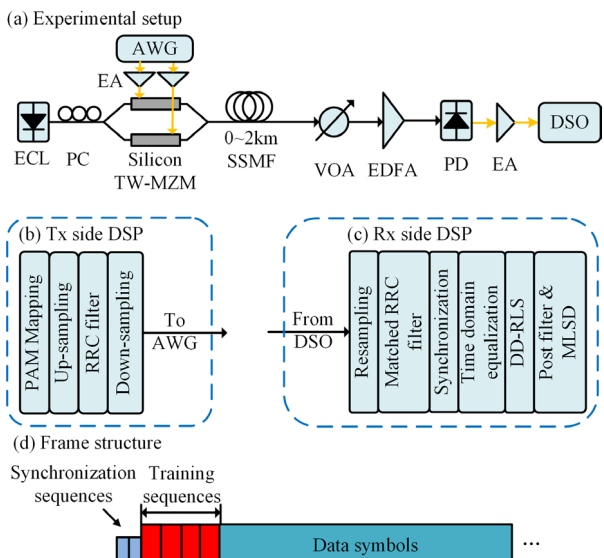

Fig. 2. (a) Experimental setup. (b) Transmitter side DSP. (c) Receiver side DSP. DD-RLS: decision directed recursive least square equalization. (d) Frame structure of baseband Nyquist PAM-4 signal.

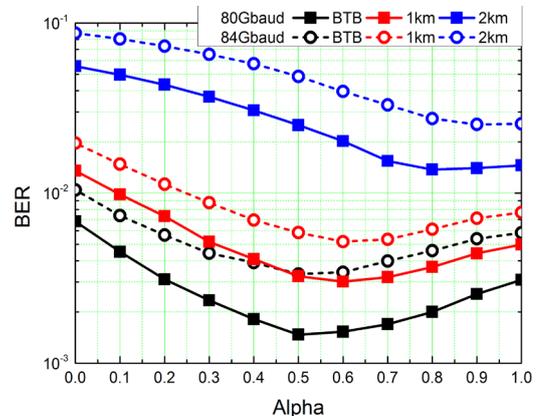

Fig. 3. Measured BER versus alpha in post filter for 80/84Gbaud PAM-4 signal at BTB scenario, after 1km and 2km SSMF transmission, respectively.

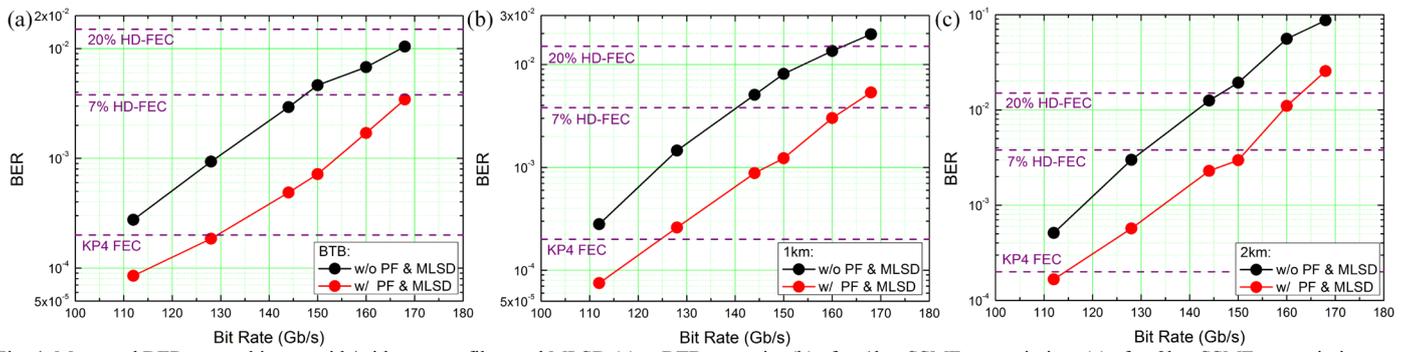
Fig. 4. Measured BER versus bit rate with/without post filter and MLSD (a) at BTB scenario; (b) after 1km SSMF transmission; (c) after 2km SSMF transmission.

case, the BER of 128Gb/s PAM-4 signal is below the KP4 FEC threshold of $2.2\times10^{-4}$ [6] with post filter and MLSD. At 7% HD-FEC threshold of $3.8\times10^{-3}$, 144Gb/s and 168Gb/s data rate can be achieved with and without post filter and MLSD, respectively. If post filter and MLSD are not applied, the BER of 168Gb/s PAM signal is still lower than the 20% HD-FEC threshold of $1.5\times10^{-2}$. For 1km SSMF transmission, 168Gb/s and 160Gb/s PAM-4 can be transmitted below the 20% and the 7% HD-FEC thresholds, respectively, by using post filter and MLSD. In addition, for 2km SSMF transmission, 160Gb/s and 150Gb/s PAM-4 signal transmission can be realized with BERs below the 20% and the 7% HD-FEC thresholds, if post filter and MLSD are applied.

Fig.5(a) shows the measured optical spectra of Nyquist PAM-4 signals at BTB scenario with different baud rates. The resolution is set as 0.02nm to see the details. Fig.5(b) depicts the measured BERs versus the received optical power for 80Gbaud PAM-4 signal at BTB scenario, after 1km and 2km SSMF transmission, respectively. It should be noted that IM signal would suffer from chromatic dispersion (CD) induced frequency-selective power fading effect with direct detection and degrade the transmission performance. To be specific, the spectrum component with frequency higher than 40GHz and 30GHz are attenuated by more than 3dB after 1km and 2km SSMF transmission, respectively. Therefore, the BERs can be below the 7% HD-FEC threshold for both BTB and 1km SSMF transmission, while the BER after 2km SSMF transmission correspond to the 20% HD-FEC threshold. At 20% HD-FEC threshold of $1.5\times10^{-2}$, compared with the BTB curve, there are 1.0dB and 5.3dB received power penalties, respectively, for 1km and 2km SSMF transmission. Fig.5(c) shows the measured BERs versus the received optical power for 84Gbaud PAM-4 signal at BTB scenario and after 1km SSMF transmission. Compared with the BTB curve, 1.2dB received power penalty can be observed at the 20% HD-FEC threshold of $1.5\times10^{-2}$ after 1km SSMF transmission.

## V. CONCLUSIONS

In this paper, we report the generation and direct detection transmission of 84Gbaud and 80Gbaud Nyquist PAM-4 signals with a silicon TW-MZM. With the help of post filter and MLSD, the measured BERs of 168Gb/s and 160Gb/s PAM-4 signals, after 1km and 2km SSMF transmission, respectively, are below the 20% HD-FEC threshold. The net data rates are 135.7Gb/s and 129.2Gb/s, respectively. To our best knowledge, we demonstrate the highest single lane data rate of PAM-4 signal transmission based on silicon TW-MZM. Our work indicates that silicon TW-MZM has great potential for low-cost data center interconnection.

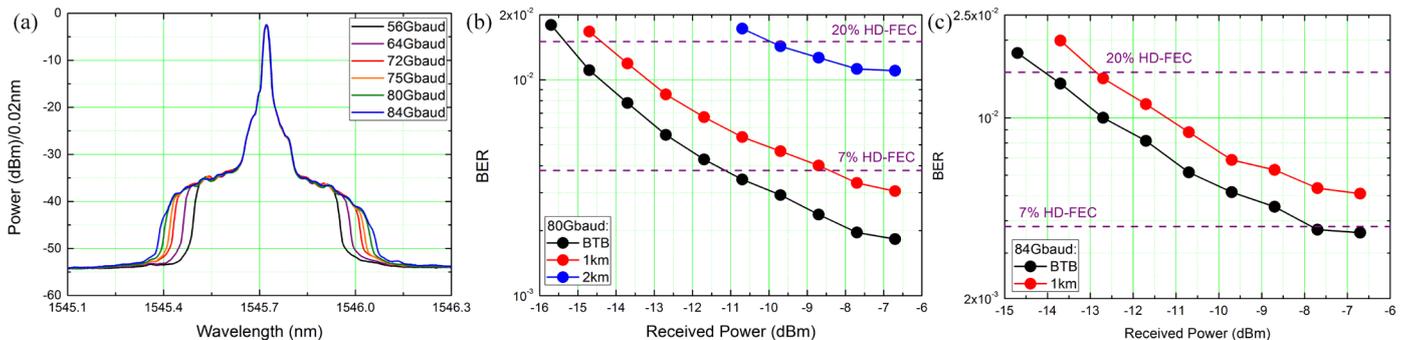
Fig. 5. (a) Measured optical spectra of Nyquist PAM-4 signals at BTB. (b) Measured received power for 80Gbaud PAM-4 signal at BTB, after 1km and 2km SSMF transmission. (c) Measured received power for 84Gbaud PAM-4 signal at BTB and after 1km SSMF transmission.